\documentclass{article}

\usepackage{arxiv}

\usepackage[utf8]{inputenc} 
\usepackage[T1]{fontenc}    
\usepackage[hyphens,spaces,obeyspaces]{url}
\usepackage{hyperref}       
\usepackage[hyphenbreaks]{breakurl}
\usepackage{booktabs}       
\usepackage{amsfonts}       
\usepackage{nicefrac}       
\usepackage{microtype}      
\usepackage{lipsum}
\usepackage{graphicx}
\usepackage{authblk}
\usepackage[bottom]{footmisc}
\graphicspath{ {./images/} }

\title{A Comparison of the Cerebras Wafer-Scale Integration Technology with Nvidia GPU-based Systems for Artificial Intelligence}

\author[ ]{Yudhishthira Kundu}
\author[ ]{Manroop Kaur}
\author[1]{Tripty Wig}
\author[1]{Kriti Kumar}
\author[1]{Pushpanjali Kumari}
\author[1]{Vivek Puri}
\author[1]{Manish Arora}

\affil[1]{Insaito, Inc., 4695 Chabot Drive \#200, Pleasanton, CA, 94588, USA\\ contact@insaito.com}

\let\OLDthebibliography\thebibliography
\renewcommand\thebibliography[1]{
  \OLDthebibliography{#1}
  \setlength{\parskip}{0pt}
  \setlength{\itemsep}{0pt plus 0.2ex}
}

\begin{document}
\maketitle
\begin{abstract}
Cerebras’ wafer-scale engine (WSE) technology merges multiple dies on a single wafer. It addresses the challenges of memory bandwidth, latency, and scalability, making it suitable for artificial intelligence. This work evaluates the WSE-3 architecture and compares it with leading GPU-based AI accelerators, notably Nvidia’s H100 and B200. The work highlights the advantages of WSE-3 in performance per watt and memory scalability and provides insights into the challenges in manufacturing, thermal management, and reliability. The results suggest that wafer-scale integration can surpass conventional architectures in several metrics, though work is required to address cost-effectiveness and long-term viability. 

\textbf{Keywords:} Wafer Scale Integration, WSE, GPU Hardware, AI Training, AI Inference
\end{abstract}


\section{Introduction}
The growing complexity of artificial intelligence (AI) and machine learning (ML) workloads has necessitated advancements in computational hardware capable of addressing memory bandwidth, latency, and scalability challenges. Traditional multichip architectures have shown interchip communication bandwidth bottlenecks and added latencies, leading to the "Memory Wall"~\cite{mckee2004reflections}~\cite{wulf1995hitting} and prompting the exploration of novel architectures. 
 
To address these challenges, Nvidia has launched a series of GPU-based AI accelerators, such as the Hopper H100~\cite{h100} and the Blackwell B200~\cite{b200}. The H100 built in the TSMC 4nm process consists of 80 billion transistors. The H100 uses HBM3 memory with a capacity of 80 GB and 3 TB/s of bandwidth available. The integration of NVLink 4 supports up to 900 GB/s interconnect in multi-GPU systems. The B200 has 208 billion transistors fabricated in the TSMC 4 nm process, providing 20 PetaFLOPS of FP4 AI performance. It gives 8 TB/s Memory Bandwidth using 8-site HBM3e memories and 1.8 TB/s of bidirectional NVLink bandwidth. 

 Another recent development by Nvidia is the integration of multiple dies within a single package to create a "Superchip." For example, the Nvidia GB200 Grace Blackwell superchip connects two B200 Tensor Core GPUs to the Grace CPU over a chip-to-chip interconnect~\cite{nvidiaNVIDIABlackwell}.
 
 This approach of integrating multiple-dies has existed for some time through the idea of "Wafer-Scale Integration (WSI)"~\cite{lea1990wasp}~\cite{pal2019architecting}. These systems integrate over a whole wafer, enabling much larger integration than a single die. For example, in 1980, Trilogy Systems~\cite{wikipediaTrilogySystems} attempted to build an IBM-compatible mainframe using wafer-scale integration, producing a single chip that was 2.5 inches on one side. More recently, Cerebras Systems has made significant advancements by developing commercially viable~\cite{lauterbach2021path}, large-scale wafer-scale processors specifically designed for AI applications. Cerebras's "Wafer Scale Engine (WSE)"~\cite{lie2021multi}~\cite{lie2023cerebras}~\cite{lie2024inside} is considered a leading example of the WSI technology.
 
 Approaches like the WSE interconnect dies on a wafer to reduce inter-chip communication delays and optimize power and performance. Die-2-die communication on a wafer inherently provides higher bandwidth and lower latency than off-die communication methods crossing transceivers and package boundaries, require additional power and introduce delays due to the physical separation between components.

 Cerebras first released a WSE product in 2019. The Wafer Scale Engine 1 (WSE-1)~\cite{cerebras2024chip} was fabricated using TSMC’s 14nm technology and demonstrated the feasibility of WSI in AI processors. With a die size of 46,225 mm², the WSE-1 contained over 1.2 trillion transistors, 400,000 AI-optimized cores, and 18 GB of high-speed SRAM. It achieved a memory bandwidth of 9 petabytes per second, enabling it to handle large-scale AI workloads by minimizing inter-chip communication and associated latency and power costs.

 The next release, the Wafer Scale Engine 2 (WSE-2)~\cite{cerebras2024wse2}, was manufactured using TSMC's 7nm technology and extended the capabilities of WSE-1. With 2.6 trillion transistors, 850,000 cores, and 40 GB of on-chip SRAM, the WSE-2 achieved a memory bandwidth of 20 petabytes per second and a fabric bandwidth of 220 petabits per second. The architecture retained the WSE-1’s single-wafer design while improving computational throughput and energy efficiency, making it suitable for replacing traditional GPU clusters in AI training and inference tasks.

 Most recently, the Wafer Scale Engine 3 (WSE-3)~\cite{cerebras2024wse3} was released in 2024. It was developed using the TSMC 5nm process and integrates 4 trillion transistors, 900,000 AI-optimized cores, and 44 GB of on-chip SRAM. It achieves a peak computing performance of 125 petaflops and a memory bandwidth of 21 petabytes per second. The WSE-3 is designed to support large language models with up to 24 trillion parameters, enabling its deployment in high-performance AI supercomputing systems.

With increased attention around AI, WSI-based processors such as the Cerebras WSE offer a powerful alternative to Nvidia's GPU-based systems~\cite{cerebrasnews1}~\cite{cerebrasnews2}. Hence it becomes important to evaluate these systems with each other and weigh their pros and cons. This paper thoroughly examines the CS-3 AI system~\cite{cerebrasCerebrasCS3} based on WSE-3 silicon. It focuses on its comparative study of architecture, computational performance, and key technological parameters with current SOTA (state-of-the-art) AI systems based on Nvidia GPUs. 

Please note that the Cerebras CS-3 is powered by the "Wafer Scale Engine 3" (WSE-3) chip, which is the core processing unit within the CS-3 system. In this paper, we may use the CS-3 and WSE-3 interchangeably.

\section{Evaluation Framework}
The CS-3 product is designed for High-Performance AI training and inference. AI workloads bring their own set of unique bottlenecks and challenges for computing systems~\cite{ren2019performance}~\cite{zhang2024llmcompass}~\cite{zhang2024benchmarking}. The main metrics impact performance are as follows. The section also explains why each of these metrics matters for AI workloads.

We evaluate the metrics below to find out how WSE fares, and also what new challenges are being exposed by this approach. However, the evaluation must be done in the context of a system and not amongst un-equals, e.g., it is not fair to compare a WSE to a single GPU. This is because the goal of building a system is to pack more compute horsepower into a fixed physical footprint with lower power consumption and cooling costs. 

A typical AI Compute architecture is hierarchical, consisting of servers and racks. Typically, each server has 1 to 8 GPUs. Hence, each 42u data center rack has about 42 to 256 GPUs. The rack represents a fixed-size real estate to hold the compute capacity, and hence, to normalize the comparison we choose the rack as a point of granularity to make the evaluation i.e., a server rack designed with H100-based GPUs vs. a server rack built out of Cerebras WSE-3. In addition to making the comparison more meaningful, the ISO rack size performance is translated to performance per watt (perf/W) ~\cite{juniper2024}.

\subsection{Memory Bandwidth}
AI and deep learning workloads typically involve moving large volumes of data—whether it’s model weights, activation maps, or feature embeddings—between memory and compute units (CPUs, GPUs, or specialized accelerators)~\cite{gholami2024ai}. The rate at which data can be transferred (memory bandwidth) often becomes the bottleneck.

\textbf{Feeding the Compute Units}: Even with powerful processors, if data cannot be supplied quickly enough, many cycles are spent idle waiting on data (a “memory-bound” scenario).

\textbf{Mini-Batch Throughput}: Training efficiency for large neural networks often requires quickly loading and storing mini-batches. Higher bandwidth allows bigger mini-batches (or more frequent updates) without stalling.

\textbf{Overall Utilization}: GPUs and specialized AI accelerators are designed for massive parallelism; high bandwidth is crucial for keeping all those parallel units busy.

\subsection{Scalability}
“Scalability” refers to a system’s ability to handle increasing workloads by leveraging additional resources. For AI, especially at enterprise or hyperscale levels, it’s crucial that adding more GPUs/nodes or more powerful hardware continues to produce meaningful gains in throughput or reduced time-to-train~\cite{supermicroOptimizingWorkloads}.

\textbf{Linear vs. Diminishing Returns}: Ideally, doubling the number of compute nodes halves the training time. In practice, communication overhead, synchronization barriers, and shared resource contention can cause less-than-linear speedups.

\textbf{Algorithmic Scalability}: Not all AI algorithms scale equally. Some require frequent global communication (e.g., large distributed model training with frequent parameter updates), which can limit scaling.

\subsection{Memory Capacity}
Memory capacity determines how much data can be stored “close” to the processor. AI/ML models—particularly large language models or image recognition systems—often have massive numbers of parameters. During training or inference, all or part of these parameters, along with intermediate data (activations), must be in memory~\cite{gholami2024ai}.

\textbf{Model Size Constraints}: If the model (or its critical parts) doesn’t fit into GPU/accelerator memory, frequent data transfers or model partitioning across multiple devices becomes necessary, adding complexity and overhead.

\textbf{Batch Size and Layer Caching}: Larger memory allows larger mini-batches and caching of intermediate layers, often improving throughput and reducing training iterations.

\textbf{Out-of-Core Computation}: If memory capacity is too small, the system may have to swap data to host memory or slower storage, drastically reducing performance.

\subsection{Scale Up and Scale Out}
When an organization wants to boost performance, it typically faces a choice between “scaling up” (using bigger, more powerful machines with more CPUs/GPUs and memory in a single node) or “scaling out” (adding more nodes in a distributed setup)~\cite{ibmVerticalScaling}.

\textbf{Scale-Up (Vertical Scaling)}: This approach involves using higher per-node memory capacity and bandwidth which reduces the latency for intra-node communication. This enables potentially simpler software stack if everything runs on a single large machine. However, these machines can be very expensive and have a practical limit in size and cost.
    
\textbf{Scale-Out (Horizontal Scaling)}: In this approach, more nodes working in parallel can tackle much larger datasets or bigger models collectively and can potentially give near-linear speedups if the workload is highly parallelizable and the network/communication overhead is well managed. However, there is more complexity in distributed training frameworks (e.g., synchronized parameter updates across many nodes) and potential bottlenecks if network bandwidth or latency cannot keep up.

\subsection{Power and Thermals}
Power consumption and heat dissipation are practical constraints in AI data centers~\cite{aipower}~\cite{mhiDataCenter}. As compute density increases, systems can quickly become limited by how much they can be cooled.

\textbf{Performance per Watt}: AI accelerators (e.g., GPUs, TPUs) are often evaluated on how many TFLOPS (teraFLOPS) they can deliver per watt. Higher efficiency means you can run more compute in the same power envelope.

\textbf{Thermal Throttling}: If cooling is insufficient, processors will reduce clock speeds to prevent overheating, which directly impacts performance.

\textbf{Data Center Design}: Large-scale AI computing clusters require advanced cooling solutions (liquid cooling, immersion cooling, or advanced airflow designs).

\subsection{Yield and Fault Tolerance}
As AI systems grow in scale, both at the chip level (e.g., large GPUs) and data center level (thousands of servers), the chance of hardware failures grows~\cite{xu2024fault}~\cite{cerebras100xDefect}. Yield refers to how many functional chips are produced in manufacturing; fault tolerance covers how a system handles failures at runtime.

\textbf{Hardware Yield}: Large AI-specific chips can be more prone to defects. Lower yields can increase costs. Manufacturers may salvage partially defective chips by disabling defective sections, which might reduce performance.
    
\textbf{Reliability in Large Clusters}: At scale, partial failures (e.g., a single GPU in a multi-node job failing) can cause job restarts unless the system has robust checkpointing and fault tolerance. Systems like distributed training can be designed for graceful handling of node drop-outs or network issues, but these features add overhead and complexity.

\section{Key Architecture Differences} 
Before we discuss the detailed evaluation results of the CS-3 and current SOTA AI systems, let's understand the key differences between WSE-3 architecture from H100/B200. Below is a concise comparison based on an analysis of the literature~\cite{lauterbach2021path}~\cite{lie2021multi}~\cite{lie2023cerebras}~\cite{lie2024inside}~\cite{cerebras2024wse2}~\cite{cerebras2024chip}.

\subsection{Decoupled Memory and Compute}
\textbf{Cerebras WSE-3}: Separates (decouples) memory from the computational cores. Parameters can live in external memory (MemoryX or similar) while compute elements on the wafer handle the processing. This design enables large model support without tying memory size directly to on‑chip GPU RAM. 

Attaches external memory modules (e.g., MemoryX) that can scale independently of wafer compute. The independent memory cluster allows independent model size scaling. Users can grow the memory cluster to handle larger models without changing the wafer.

\textbf{Nvidia H100/B200}: Uses an integrated GPU memory system—each GPU has onboard HBM memory closely coupled with its cores.

The model size is limited by the GPU’s onboard HBM memory or the capacity of distributed GPU clusters, which must be carefully partitioned and synchronized across multiple GPUs.

\subsection{Data-Level Parallelism}
\textbf{Cerebras WSE-3}: Focuses on data parallelism across the wafer. Each core on the wafer processes a slice of data simultaneously. The architecture is streamlined for matrix and tensor operations in a highly parallel fashion. Hence the compute is pure Data-Level Parallelism.

\textbf{Nvidia H100/B200}: Also supports data-level parallelism but often combines it with pipeline parallelism and other scheduling strategies.

\subsection{Execution Flow}
\textbf{Cerebras WSE-3}: Implements a “layer-by-layer” execution flow and one layer of the learning network is typically processed at a time. The entire wafer is devoted to computing one layer for all data, then moves on to the next layer. This avoids typical GPU overheads in scheduling and memory synchronization across multiple layers concurrently.

\textbf{Nvidia H100/B200}: Can concurrently process multiple layers or parts of layers (e.g., pipeline parallelism), often requiring more complex synchronization and partitioning.

\subsection{Storage of Weights}
\textbf{Cerebras WSE-3}: Weights are stored and recalculated from Backpropagation. The system stores weights in external memory and streams them onto the wafer for forward and backward passes. Gradients and updated weights can be efficiently recalculated because the architecture is designed for rapid parameter streaming.

\textbf{Nvidia H100/B200}: Typically stores weights in GPU memory. Data transfers can become a bottleneck if the model exceeds on-board memory, requiring partitioning across multiple GPUs and more complex communications.

\subsection{Die-to-Die Communication}
\textbf{Cerebras WSE-3}: Implements on‑wafer, die‑to‑die communication across its massive 2D mesh of compute cores. This high-bandwidth, ultra-low-latency interconnect is part of the wafer-scale design.

\textbf{Nvidia H100/B200}: Relies on NVLink or PCIe for GPU-to-GPU or GPU-to-CPU communication. While NVLink is high-bandwidth relative to older interconnects, it remains off-die and thus inherently higher latency compared to on-wafer networks.

\subsection{Reduced Latencies}
\textbf{Cerebras WSE-3}: Data movement latency is drastically lower by integrating compute cores and communication fabric within a single wafer-scale chip. The large on-die mesh minimizes overhead for cross-core communication. 

\textbf{Nvidia H100/B200}: Though high-performance, GPUs still contend with off-chip communication latencies—either via NVLink, PCIe, or networking in multi-GPU systems.

\subsection{Summary}

To summarize the key differences:
\begin{enumerate}
\item Layer-by-layer execution and decoupled memory allow Cerebras’s wafer-scale engine to handle huge models without the usual GPU memory constraints.

\item The on-wafer, die‑to‑die interconnect bypasses many latency bottlenecks of multi-GPU setups, streamlining data movement.

\item Pure data-level parallelism across thousands of cores on the wafer enables a simpler programming model (one layer at a time) rather than juggling pipeline- or model-parallel strategies often needed on GPUs.
\end{enumerate}

Overall, Cerebras's WSE-3 is architected to minimize data communication overhead and maximize usable memory for ever-larger models, in contrast to Nvidia’s more traditional (though high-performance) GPU-based design that couples compute and memory on each GPU and relies on external interconnects between GPUs.

\renewcommand{\arraystretch}{1.2}

\begin{table}[h]
    \centering
    \small
    \setlength{\tabcolsep}{5pt}
    \caption{Rack ISO Space - CS-3, DGX H100, and DGX B200 Components}
    \label{tab:rack-iso-space}
    \begin{tabular}{l l l l}
        \toprule
        \textbf{Component} & \textbf{WSE-3} & \textbf{H100} & \textbf{B200} \\
        \midrule
        \textbf{Chip Size} & 46,225 mm\(^2\) & 814 mm\(^2\) & \(\sim\)1600 mm\(^2\) \\
        \textbf{\# Cores/Chip} & 900000 & 16896 FP32 & - \\
        \textbf{On-Chip Memory/H100} & 44 GB & 0.05 GB & - \\
        \midrule
        \textbf{System} & CS-3 & DGX H100 & DGX B200 \\
        \midrule
        \textbf{System Dimension} & 15U & 8U & 10U \\
        \textbf{\# Chips/System} & 1 & 8 & 8 \\
        \textbf{On-Chip Memory/H100} & 44 GB & 0.4 GB & - \\
        \textbf{Memory Capacity} & 1.2-1,200 TB & 0.64 TB & 1.5 TB \\
        \textbf{System Power} & 23 kW & 10.4 kW & 14.3 kW \\
        \textbf{Price} & \$2.5M (est.) & \$0.35M & \$0.5M \\
        \midrule
        \textbf{Rack Dimension: ISO Space} & 30-32U & 30U & 32U \\
        \midrule
        \textbf{\# Systems/Rack} & 2 & 4 & 3 \\
        \textbf{\# Chips/Rack} & 2 & 32 & 24 \\
        \textbf{On-Chip Memory} & 44 GB & 1.6 GB & - \\
        \textbf{Memory Capacity} & 1.2-1,200 TB & 2.56 TB & 4.5 TB \\
        \textbf{\# Cores/Rack} & 900000 & 33792 FP32 & - \\
        \textbf{Rack Price} & \$5M (est.) & \$1.4M & \$1.5M \\
        \textbf{Rack Power} & 46 kW & 41.6 kW & 43.9 kW \\
        \bottomrule
    \end{tabular}
\end{table}

\section {Evaluation}

This section presents a detailed comparison of the CS-3 vs. Nvidia H100/B200. We will compare the systems on raw performance and evaluate them on other key technology factors such as Scalability, Yield, Packaging and Assembly, Power Delivery, etc.

\subsection {Raw Performance}
For comparison purposes, both H100 and CS-3 systems are normalized for ISO space and ISO power~\cite{ exxact2024}~\cite{nvidia2024}~\cite{anandtechNVIDIABlackwell}. Table~\ref{tab:rack-iso-space} summarizes the capacity enabled using of the respective systems for a 30--32U rack system configuration. A single CS-3 fits within 15U space, whereas 8 H100 and B200 GPUs can fit within an 8U--10U volume in the form of a single DGX system~\cite{nvidia2024}. The WS-3 can provide a much larger memory capacity as it uses external memory. The H100 integrates 80GB of HBM memory per GPU with the B200 expected to increase that number by 2x--3x. The price for each CS-3 is estimated to be \$2M to \$3M~\cite{datacenterdynamicsCerebrasUnveils}, giving it a much higher rack price than Nvidia-based systems at current prices. Each CS-3 server is said to consume 23KW~\cite{datacenterdynamicsCerebrasUnveils}. That brings the rack power for all of the systems designs to be between 41kW--46kW.

Table~\ref{tab:performance} compares the peak performance for the CS-3 vs. H100 and B200 systems for an ISO space and ISO power rack. The B200 is expected to provide 3x--4x performance as compared to the H100~\cite{exxactcorpComparingBlackwell}. The CS-3 system delivers notable performance advantages~\cite{cerebras2024comparison}. As seen in Table~\ref{tab:performance}, as compared to the H100-based system, the CS-3 achieves approximately 3.5x FP8 AI peak-performance and 7x FP16 AI peak-performance. When compared to the B200, the advantage drops, but the CS-3 still delivers about 1.1x FP8 performance and 2.15x FP16 performance. However, when considering performance normalized for ISO-space, power, and cost (performance/watt/\$), the B200 system offers significant advantages over CS-3. The B200 delivers about 1.5x--3x better metrics.

\renewcommand{\arraystretch}{1.2}

\begin{table}[h!]
    \centering
    \caption{Performance Comparison for CS-3, H100, and B200. The FP8 and FP16 numbers are peak FLOPS in PetaFLOPS.}
    \label{tab:performance}
    \begin{tabular}{@{} l l l l l l l l @{}}
        \toprule
        \textbf{System}  & \textbf{FP8}  & \textbf{FP16} & \textbf{Power (kW)}  & \textbf{FP8/W}  & \textbf{FP8/W/\$} & \textbf{FP16/W} & \textbf{FP16/W/\$} \\ 
        \midrule
        CS-3            & 250          & 250           & 46              & 5.43           & 1.09             & 5.43           & 1.09             \\
        H100           & 64           & 32            & 41.6            & 1.54           & 1.10             & 0.77           & 0.55             \\
        B200           & 216          & 108           & 42.9            & 5.03           & 3.35             & 2.52           & 1.68             \\ 
        \midrule
        \textbf{CS-3 Advantage} \\ 
        vs. H100       & 3.9x         & 7.8x          & -               & 3.52x          & 1.00x            & 7.05x          & 2.00x            \\
        vs. B200       & 1.16x        & 2.31x         & -               & 1.08x          & 0.32x            & 2.15x          & 0.65x            \\ 
        \bottomrule
    \end{tabular}
\end{table}

\subsection{Scalability}
In AI systems, scalability is needed for both memory/storage and compute to increase capability and compute. In Nvidia systems, compute and memory are directly linked, so as you scale the compute, the memory increases, and vice versa. However, the compute-to-memory ratio is fixed for a given silicon/system. For a system, if only more memory is needed, compute also needs to be increased as they are tightly coupled~\cite{nvidia2024gpumemory}. 

However, in WSE-3 architecture, the compute and memory are decoupled, and the memory capacity of a single CS-3 scales from 1.2 TB to 1,200 TB. It is known that the memory required for training is approximately 8GB-16GB for every 1 billion model parameters~\cite{Thiyagu2024}. Also, if the AI model is restricted to a single chip, it will have significant latency advantages. The time to first token will also be better as moving across chips adds latency and increases power consumption~\cite{Thiyagu2024}. The fact that the CS-3 can connect to much larger amounts of memory allows it to train larger models.

\renewcommand{\arraystretch}{1.2}

\begin{table}[h!]
    \centering
    \caption{Model Size vs Compute and Memory Capacity}
    \label{tab:model-size-compute}
    \begin{tabular}{@{} l l l l l l @{}}
        \toprule
        \textbf{Parameters (Billion)} & \textbf{Memory (GB)} & \textbf{B200 Chips (\#)} & \textbf{B200 DGX (\#)} & \textbf{WSE-3 (\#)} & \textbf{WSE-3 Memory (TB)} \\ 
        \midrule
        10      & 80      & 1   & 1   & 1   & 1.2  \\
        100     & 800     & 5   & 1   & 1   & 1.2  \\
        1,000   & 8,000   & 42  & 6   & 1   & 8.4  \\
        10,000  & 80,000  & 417 & 53  & 1   & 80.4 \\ 
        \bottomrule
    \end{tabular}
\end{table}

Table~\ref{tab:model-size-compute} explains the number of systems required to train AI models of different sizes, from 10 billion parameters to 10 trillion parameters. The memory column lists the memory requirements, assuming 8GB of memory is needed to train for every 1 billion parameters. The following two columns list the number of B200 chips and DGX systems (each DBX is 4 B200 chips) needed for the necessary memory capacity. For each model size, only 1 WSE-3 is needed; last column lists the amount of memory that needs to be connected to the WSE-3. The table does not account for the compute requirements to train the model in a certain fixed time.

In summary, CS-3 has significant advantages in scalability as memory capacity can be increased without increasing compute. Each CS-3 can pack significantly more computer power as it has about 50x more area than a typical GPU. It can avoid/minimize chip-2-chip communication, improving latency and reducing the time to first token (TTFT) response time~\cite{cerebras2024comparison}. TTFT is an important metric in AI inference as it correlates with the responsiveness of the system.


As evident from the performance and scalability sections, Cerebras CS-3 has advantages over current GPU-based architectures in performance and scalability. However, its larger chip size may create many system-level challenges, adding cost and increasing the product's price, not evaluating each dimension. 

\subsection{Die-2-Die Connections}
In a standard process, die-to-die connections cross package boundaries or are handled within a package with multi-die integration in a package. In such cases, the packaging is the central construct used to create die-2-die connections.

The standard fabrication process is one of step and repeat, which produces identical independent die on the wafer and leaves scribe lines between them. The scribe lines are where the wafer is cut to create separate dies that are packaged within chips. Fabs also place test and control structures for the fab process within the scribe line spaces between the die.

Working with TSMC, Cerebras has repurposed the Scribe lines as wires that connect with another die~\cite{dylanpatel}. This has enabled them to create a Wafer die-2-die connection, which significantly reduced the latency. Performance is much higher with very low power consumption, as distances are very short and transceivers are needed to move data from one die to another~\cite{CerebrasDeepDive2023}. 

This on-wafer die-2-die communication created by repurposing scribe lines provides a way to connect cores on adjacent dies with the same bandwidth as cores on the same die. It provides a very high bandwidth interface with lower latency and power consumption, as it does not need to go through package boundaries. This approach is significantly better than chip-to-chip interconnect techniques like Nvlink, albeit very specific to Cerebras.

\subsection{Yield} 
The WSE-3 has a die area of 462 cm², resulting in a higher likelihood of defects than smaller semiconductor dies, where yields typically exceed 90\%. NVIDIA GPUs achieve higher yields through an optimized manufacturing process that utilizes smaller dies, allowing for more efficient wafer usage and reasonable yields despite process complexity. Let us understand how WSE-3 handles and manages yield.

In today’s system, yield is a function of chip size and defect tolerance. The smaller the core size, the better the system will be fault tolerant. Additionally, designs implement redundancy-based defect tolerance across components such as cores, fabric, and memory to improve yields. In WSE-3, each core is 0.05mm$^2$, while in H100, the SM core is ~6mm$^2$. If there is a defect in the core area, the full core area needs to be disabled \cite{anysilicon2024}. Since each core is much smaller in the WS-3, less area may be wasted per defect.

\renewcommand{\arraystretch}{1.2}

\begin{table}[h!]
    \centering
    \caption{H100 vs WSE-3 Yield Calculations. The wafer size is 300mm at TSMC 5nm.}
    \label{tab:h100-vs-wse3}
    \begin{tabular}{@{} l l l @{}}
        \toprule
        \textbf{Metric} & \textbf{H100} & \textbf{WSE-3} \\ 
        \midrule
        Chips/Wafer                        & 72      & 1     \\
        Total Chip Area (mm$^2$)           & 58,608  & 46,225 \\
        Defect Rates (mm$^{-2}$)           & 0.001   & 0.001  \\
        Total Defects                      & 59      & 46     \\
        Fault Tolerant Core Size (mm$^2$)  & 6.2     & 0.05   \\
        Maximum Die Space Lost (mm$^2$)    & 361     & 2.2    \\ 
        \bottomrule
    \end{tabular}
\end{table}

Table~\ref{tab:h100-vs-wse3} shows the calculations for the maximum die size that could be lost to defects. Counter-intuitively, even though the WSE-3 is a much larger chip, the die space lost is lower. This is because each defect impacts a lesser area on the WSE-3. Hence, the WSE-3 is about 164x more fault tolerant for cores than the H100. 

However, the chip does not have only cores; about 50\% are used by SRAMs, register files and on-chip fabrics. These components can be designed with redundancy to recover yields. As an example, the WSE-3 has designed the dynamic configurable on-chip fabric to dynamically change connections between cores. This helps recover almost full yield. 

In summary, WSE-3 has addressed the yield challenges of bigger silicons due to defect densities by designing very small processing cores with the flexibility of dynamically configurable fabric and other redundancy techniques. The yield of WSE-3 is expected to be in the same ballpark as reticle-limited die sizes~\cite{anysilicon2024}.

\subsection{Packaging and Assembly} 
Cerebras has significantly pushed the state-of-the-art and solved key engineering challenges to get the system to work at scale from a packaging and assembly perspective.

In the WSE-3 a full wafer works as a  chip ~46,225mm$^2$ while the max area of a traditional chip is about 815 mm$^2$ due to reticle limits. Hence it seems that the traditional assembly and packaging techniques will pose challenges~\cite{cerebras2024wafer}. 

In manufacturing such big packages, there are three main challenges from a packaging and assembly perspective. First is the cost. This is reflected in the high cost of a CS-3 system. Second, power delivery needs to be provisioned to deliver more than 20,000 amperes of currents with very good voltage regulation. This can contribute significantly to the cost. Lastly, thermal dissipation is a challenge, and the system needs to be designed to extract about 20kW of heat from the wafer.

Traditional packaging methods are not designed to handle such scale and will not work directly. A lot of innovation and experimentation is needed to package the WSE-3 and feed the required power with efficient cooling~\cite{ieee2021wafer}. As an example, the WSE-3 is “packaged on board” instead of traditional packaging. That means that the wafer is directly mounted on the board. This eliminates the need of a separate package, provides a smaller footprint, improves SI (Signal Integrity), and reduces cost. In short, PCB becomes the “package” itself. 

To make this work, the WSE-3 employs a multi-layered assembly consisting of a printed circuit board (PCB), a flexible membrane, the WSE-3 die, and a heat exchanger. This structural arrangement supports mechanical stability and heat dissipation accommodating the wafer's thermal expansion and contraction~\cite{cerebras2024wafer}. 

In assembly, the fact that materials respond differently to heat presents a fundamental issue. The wafer-scale chip is mounted on the PCB. However, the silicon and PCB materials expand at different rates under changes in temperature. Thus, things aligned when the system is cool may get slightly displaced when it heats up. 

The largest displacement occurs at the edges of the connection between the silicon chip and the PCB. In a smaller chip, this displacement is small enough that the chip to PCB connections (wires) can flex slightly and still work. However at the size of the wafer, the differences in expansion between the two materials would stress these connections enough to break some if traditional packing techniques were used~\cite{Cadence2024}. 

The packaging design also uses a layered configuration that secures the wafer between the PCB and the heat exchanger with clamping fasteners. This arrangement distributes mechanical force across the assembly, maintaining electrical contacts and structural integrity. 

With all traditional attachment techniques foreclosed due to thermal mismatch, Cerebras invented a new material and designed a connector. This custom connector mates the wafer to the main PCB while absorbing the thermal displacement without breaking any electrical connections.

This sandwich of wafer, connector, and main PCB must be packaged with a fourth component, a cold plate that maintains the wafer temperature at a level comfortable for the electronics despite an overall power delivery in the mid-teen kilowatts range. There is no existing package that can maintain thermal and electrical contact and tolerate variable expansion in three dimensions for a system of this size. And there is no packaging machinery that can assemble one. 

To build the package, the four components must be fitted together to achieve precise alignment and then held in place with techniques that maintain that alignment through multiple power and thermal cycles. Cerebras invented custom machinery, tools, fixtures, and process software that make this all possible.

The system incorporates over 300 voltage regulation modules (VRMs) distributed across the wafer’s surface. These VRMs deliver current perpendicular to the wafer to support power distribution. Using multiple VRMs within each reticle provides redundancy in the power delivery system and allows independent regulation of power domains for each reticle~\cite{cerebras2024wafer}.

\subsection{Power Delivery}
With die sizes less than the reticle limit of approximately 800 mm$^2$, the primary method used is edge-based power delivery. However, for a chip of size greater than 46,000 mm$^2$, the traditional method will not work as the resistance in the interconnects would result in significant voltage drops, particularly at the center of the wafer, leading to performance inconsistencies.

Cerebras implemented a vertical power delivery system to resolve this, supplying power from above the wafer. This system works with a water-cooled cold plate beneath the chip, which dissipates heat generated by the high-density core array. The vertical approach ensures consistent voltage levels across all 900,000 cores, maintaining reliable performance during computationally intensive workloads. This solution integrates power delivery and thermal management, addressing the specific requirements of the WSE-3’s wafer-scale design~\cite{semianalysis2022die}. 

\subsection{Thermal Design}
The thermal design of WSE-3 is critical as it has a large surface area and high power consumption. Cerebras WSE-3 uses a thermal design tailored for wafer-scale systems, requiring advanced cooling solutions to manage heat across the entire wafer. A custom cold plate and connector are integrated, as even the slightest inefficiencies can lead to thermal hotspots, which may degrade the performance or lead to localized overheating. The heat is removed from the wafer using a water-cooling system. 

Water flows through micro-fins on the backside of a copper heat exchanger. The wafer is allowed to expand and contract while still in contact with the polished front side of the exchanger. This design ensures that the wafer remains thermally connected to the heat exchanger despite the differing coefficients of thermal expansion between copper and silicon. In comparison, Nvidia GPUs employ thermal designs based on modular GPU architectures, with heat dissipation managed through fans, heatsinks, and liquid cooling in some configurations~\cite{SynopsysCerebras2024}. 

\section{Conclusion}
Creating a wafer-as-a-chip through the WSE-3 is a brave and bold effort. The approach has been tried in the past. In the 1980’s, Trilogy Systems~\cite{wikipediaTrilogySystems} co-founded by Gene Amdahl~\cite{wikipediaGeneAmdahl} encountered several obstacles that could not be solved then. The Cerebras team has put engineering at work to solve problems such as yield, power delivery, thermal dissipation and assembly and packaging. At the first order these problems seem to be solved, however, it increases the cost of the system significantly as there are special materials, techniques and tools developed to address these. The long-term reliability of these solutions needs to be looked into, as these pose a risk to overall system reliability.

Architecturally, in WSE-3 there are a lot of innovations that help improve AI performance. These include the use of on-wafer, die-2-die communication to reduce latency, provide more bandwidth, and reduce power as compared to current SOTA architectures. Secondly, the decoupling of compute and memory provides easy scalability to larger model sizes supported by increased memory capacity only; in turn, this may lower cost for very large model training.

The ISO space performance/watt numbers of the CS-3-based systems is better than B200 based systems. However, the ISO space performance/watt/\$ performance is much compared to B200 systems and it is evident from the above discussions that the higher cost of solving problems associated with wafer-scale chips are contributing to it. However, systems costs may be reduced through future scale.

\bibliographystyle{abbrv}

{\footnotesize
\section*{Trademark Attribution}
Product names used in this publication are for identification purposes only and may be trademarks of their respective companies. 

\section*{Custom Research}
Insaito, Inc. works with companies to conduct custom research and generate reports. Please email us at contact@insaito.com.

\bibliography{references}}
\end{document}